\documentclass[12pt]{iopart}

\usepackage{iopams}  

\usepackage[dvips]{color}

\usepackage{graphicx}

\def\mrm{\mathrm}

\def\etal{{\it et al. }}

\def\mrm{\mathrm}

\def\s0{s_0}
\def\nubar{\overline{\nu}}
\def\pbar{\overline{p}}
\def\sratio{\langle s_2/s_1 \rangle}

\begin{document}

\title[]
{Reply to Comment on 
`Monte-Carlo simulation study 
of the two-stage percolation transition 
in enhanced binary trees'}

\author{Tomoaki Nogawa$^1$ and Takehisa Hasegawa$^2$}
\address{
$^1$ Department of Applied Physics, 
The University of Tokyo, 7-3-1 Hongo, Bunkyo-ku, Tokyo 113-8656, Japan
\\
$^2$ Graduate School of Information Science and Technology, 
The University of Tokyo, 7-3-1 Hongo, Bunkyo-ku, Tokyo 113-8656, Japan}
\ead{nogawa@serow.t.u-tokyo.ac.jp}

\begin{abstract}
We discuss the nature of the two-stage percolation transition 
on the enhanced binary tree 
in order to explain the disagreement 
in the estimation of the second transition probability 
between the one in our recent paper ( 
J. Phys. A:Math. Theor. {\bf 42} (2009) 145001)  
and the one in the comment to it from Baek, Minnhagen and Kim. 
We point out some reasons that 
the finite size scaling analysis used by them is not proper 
for the enhanced tree due to its nonamenable nature, 
which is verified by some numerical results. 
\end{abstract}


\maketitle

We have recently reported a numerical study of 
the two-stage bond percolation transition 
on the enhanced binary tree (EBT)\cite{Nogawa09}. 
Two percolation thresholds, 
$p_{c1} \approx 0.304$ and $p_{c2} \approx 0.56$, 
which respectively correspond to the divergence 
of the correlation mass and the correlation length, are obtained. 
The value of $p_{c2}$ estimated from the fractal exponent $\psi(p)$ 
is consistent with the duality relation \cite{Benjamini00}, 
$p_{c2}=1-\pbar_{c1}$, 
where $\pbar_{c1} \approx 0.436$ is the lower threshold probability 
of the dual lattice of the EBT. 
On the other hand, 
Baek, Minnhagen and Kim estimated $p_{c2} \approx 0.48$ 
for the same model 
based on the finite size scaling (FSS) analysis \cite{Baek09b}. 
This value is significantly smaller than our estimation 
while their estimation of $p_{c1}$ and $\pbar_{c1}$ 
is consistent with ours. 
Thus they concluded that the duality relation 
does not hold for the EBT 
but inequality $p_{c2} < 1 - \pbar_{c1}$ is true. 
In this article, we compare these two estimations. 
In the following we use $p_b$ to note $p_{c2} \approx 0.48$ 
obtained in \cite{Baek09b} for the distinction.

\begin{figure}[t]
\begin{center}
\includegraphics[trim=50 230 160 -240,scale=0.30,clip]{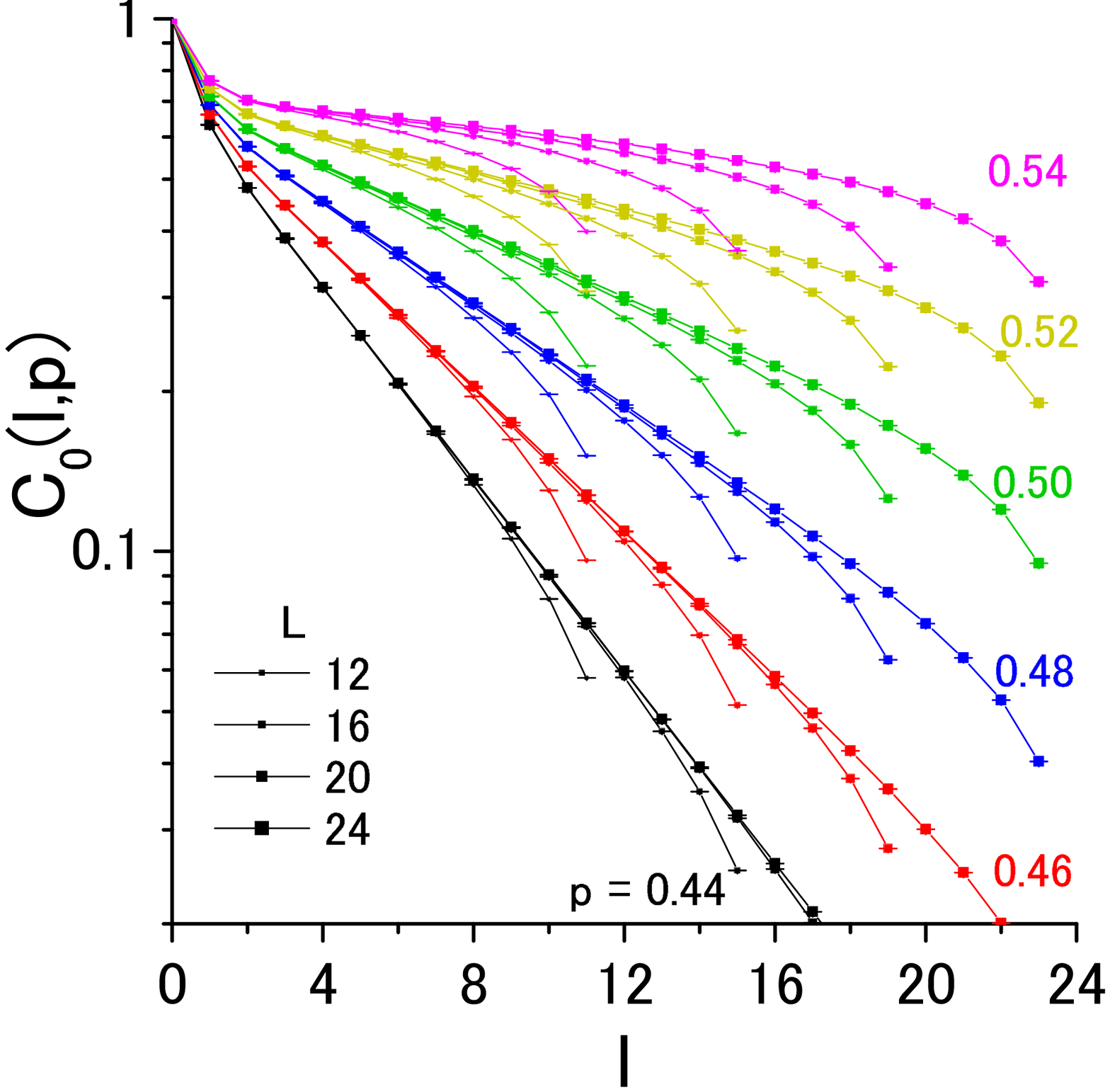}
\hspace{0.5cm}
\includegraphics[trim=50 230 160 -240,scale=0.30,clip]{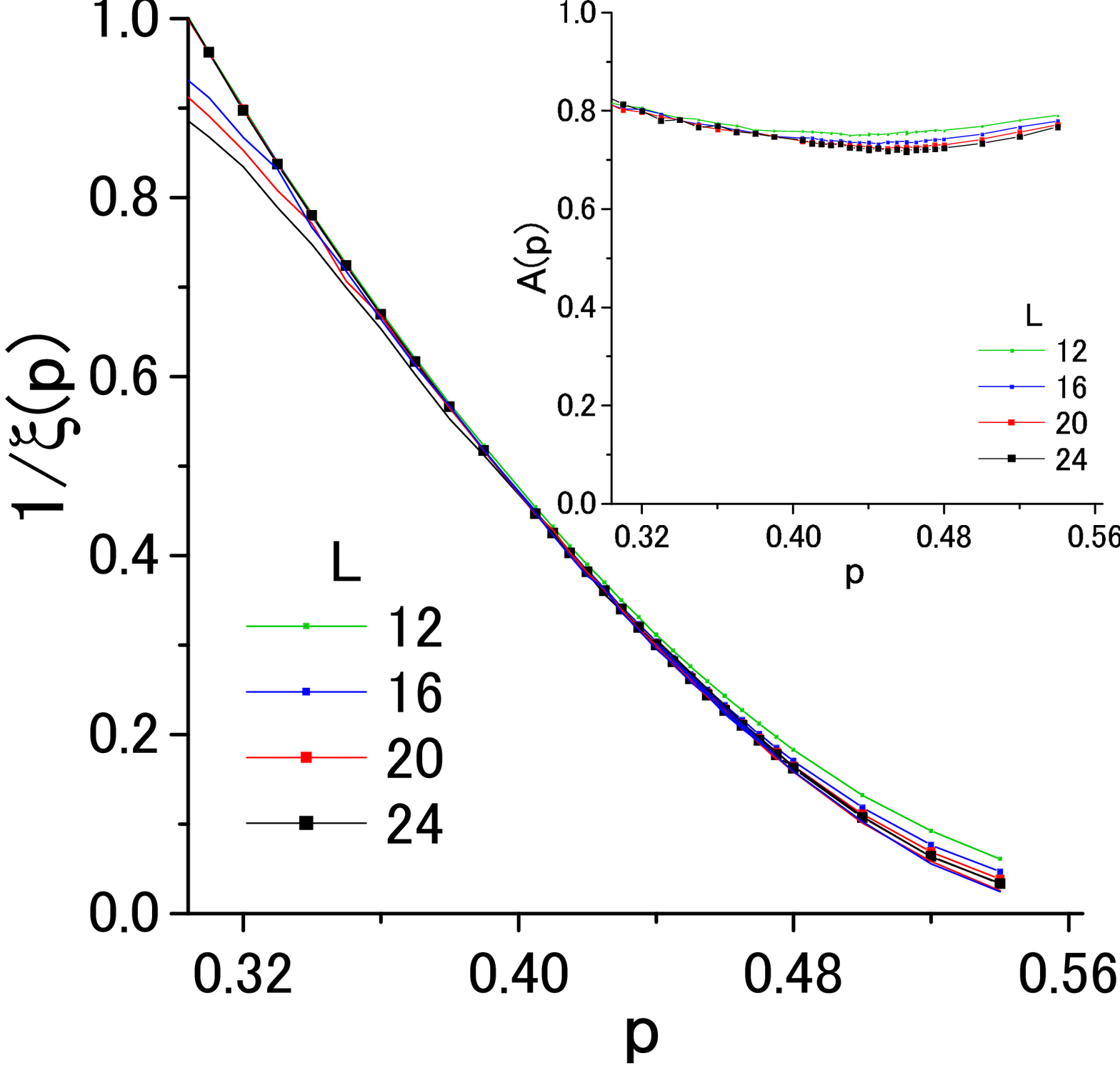}
\end{center}
\vspace{-3mm}
\caption{\label{fig:cr}
(left) The connectedness function for six $p$'s and four $L$'s. 
Exponential decay can be observed before the boundary effect appears. 
(right) $p$-dependence of (the inverse of) the correlation length. 
Symbols indicate the values calculated by 
$\xi(p,L) = -\log_2[ C_0(3L/4,p)/C_0(L/4,p)]/[L/2]$ 
and dotted lines indicate the values 
calculated from $1-\psi(p)$ \cite{Nogawa09}. 
The two estimation is almost same but 
the former is better near the $p_{c1}$ 
to reproduce $\xi(p_{c1})=1$. 
$\xi$ does not shows any singularity around $p=0.48$ 
but approaches to zero at $p \approx 0.56$.
(inset) The amplitude,  
$A(p,L) = C_0(L/2,p)/2^{-L/2\xi(p,L)}$, 
which hardly depends on $p$. 
}
\end{figure}

First, we introduce the scenario of the second transition in the EBT, 
which has been already shown in \cite{Nogawa09}. 
We only assume that connectedness function, $C_0(\ell,p)$, 
which is the probability that a site at the $\ell$-th 
generation belongs to the same cluster 
with the root site, i.e., the site at $0$-th generation, 
belongs to 
decays as a single exponential function; 
\begin{equation}
C_0(\ell,p) = A(p) 2^{-\ell/\xi(p)}= A(p) 2 ^{(\psi(p)-1)\ell}, 
\label{eq:C0}
\end{equation}
at open bond probability $p>p_{c1}$. 
Here $\xi(p)$ is a correlation length 
and $\psi(p) \equiv 1 - 1/\xi(p)$ 
is a fractal exponent of the divergent clusters. 
We confirm the exponential decay of $C_0(\ell,p)$ 
in Fig.~\ref{fig:cr}. 
Here we remarks on two quantities to detect the second transition, 
\begin{equation}
\s0(p,L) \equiv \sum_{\ell=0}^{L-1} 2^\ell C_0(\ell,p) 
\quad \mrm{and}\quad
b(p,L) \equiv 2^{L-1} C_0(L-1,p), 
\label{eq:b1}
\end{equation}
where $L$ is a number of generations of finite size samples. 
We approximately identify $x^L-1$ with $x^L$ for $x>1$ 
in the following, 
e.g., total number of nodes, $N=2^L-1 \rightarrow 2^L$. 
Substitution of eq.~(\ref{eq:C0}) into eq.~(\ref{eq:b1}) yields 
\begin{equation}
\s0(p,L) = \frac{A(p)}{2^{\psi(p)}-1} N^{\psi(p)}
\quad \mrm{and} \quad
b(p,L) = \frac{A(p)}{2^{\psi(p)}} N^{\psi(p)}.
\label{eq:b2}
\end{equation}
In these expressions, 
$b(p,L)$ and $\s0(p,L)$ are basically same quantities 
except unimportant coefficients 
and then we only treat $b(p,L)$ in the following. 
Equation~(\ref{eq:b2}) leads to an important consequence 
that $b$ is always infinite in the large size limit, 
$N \rightarrow \infty$, for $p>p_{c1}$ 
\footnote{
The first threshold is defined by 
$\xi(p_{c1})=1$ and then $\psi(p_{c1})=0$. 
}. 
Divergence of $\xi(p)$ at $p_{c2}$, 
which is indicated in the right panel of Fig.~\ref{fig:cr}, 
results that $\psi(p)$ continuously approaches to unity 
to produce an $O(N)$ term
\footnote{
Prefactor $\ell^{-\eta}$ on $C_0$ is possible but 
only results a correction factor $(\log N)^{-\eta}$ 
to $s_0$ and $b$. 
}. 
What happens at $p_{c2}$ is essentially different 
from the ordinary second order transitions in amenable graphs.

\begin{figure}[t]
\begin{center}
\includegraphics[trim=50 230 160 -240,scale=0.30,clip]{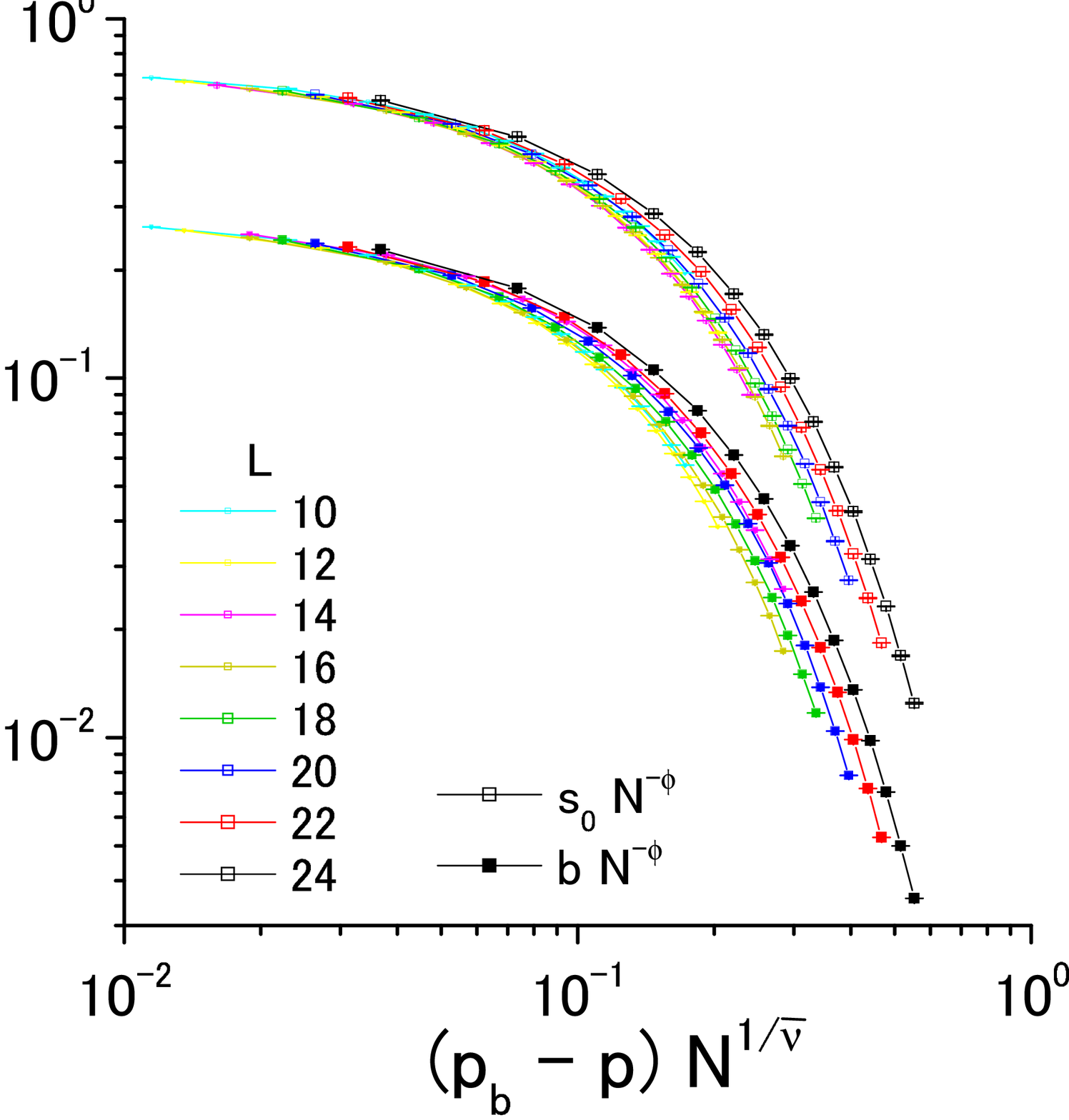}
\hspace{0.5cm}
\includegraphics[trim=50 230 160 -240,scale=0.30,clip]{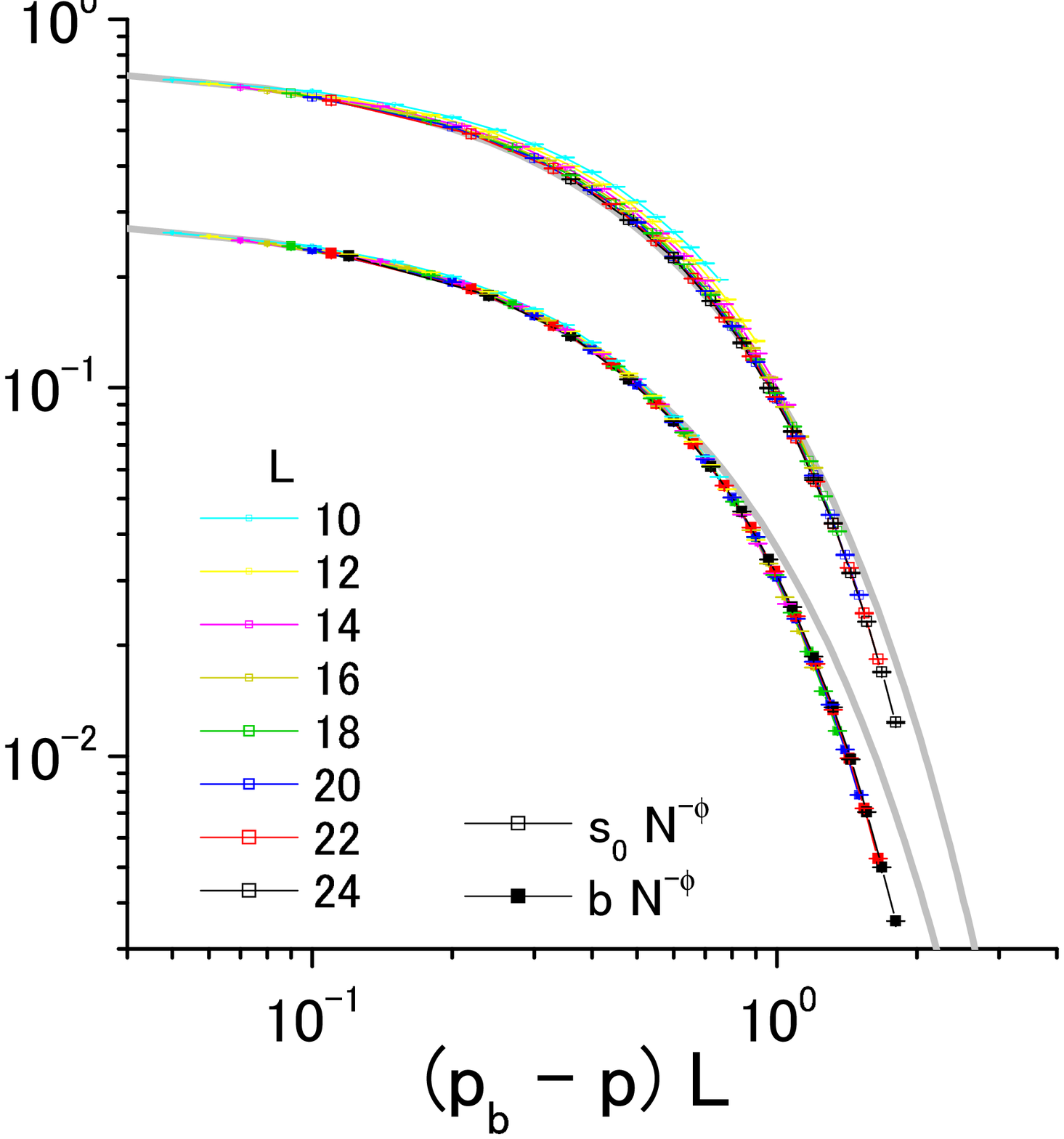}
\end{center}
\vspace{-3mm}
\caption{\label{fig:fss}
(left)  
Finite size scaling (FSS) corresponding 
to eq.~(\ref{eq:fss1}) using the parameters 
shown in \cite{Baek09b}; 
$p_b=0.48$, $\phi=0.84$ and $1/\nubar=0.12$. 
(right) 
FSS corresponding to eq.~(\ref{eq:fss2})
using $p_b=0.48$, $\phi=0.84$. 
We show guide lines proportional to $2^{-3.0 (p-p_b)L}$ 
with light gray color. 
In both scalings, 
we use the Monte-Carlo data for $0.405 < p < 0.475$ ( 0.005 step ) 
averaged with 160000 samples. 
We show the same FSS of $s_b$ together. 
}
\end{figure}

Next, we examine the analysis of Baek \etal in \cite{Baek09b}. 
They assumed a FSS formula 
\begin{equation}
b(p,L)
\propto N^{\phi} \tilde{f}_3 
\left( (p-p_b) N^{1/\nubar} \right). 
\label{eq:fss1}
\end{equation}
This formula implies, in a sense of a standard FSS, 
that $b$ is finite below $p_b$ 
and diverges as $(p_b - p)^{-\phi \nubar}$ with infinite $N$. 
This seems strange because $b$ has already diverged above $p_{c1}(<p_b)$. 
Another diverging finite component 
which results a subleading term in $b$ seems impossible 
since finite clusters growing with $p$ must be 
absorbed to the already divergent clusters before diverges by themselves. 
We consider the scaling behavior is an artifact because
eq.~(\ref{eq:fss1}) is approximately reproduced 
from eq.~(\ref{eq:b2}) without assuming another diverging component. 
Equation~(\ref{eq:b2}) leads to 
$
b(p,L)/N^{\phi}
\propto 2^{\left(\psi(p_b)-\phi \right)L 
+ \psi'(p_b)(p-p_b)L 
+ \cdots}.
$
If one chooses $p_b$ and $\phi$ satisfying $\phi = \psi(p_b)$, 
$b(p,L)/N^{\phi}$ looks a function of $(p-p_b)L$ for $|p-p_b|\ll 1$ as 
\begin{equation}
b(p,L)
\propto N^{\phi} \tilde{g}_3 
\left( (p-p_b) L \right).
\label{eq:fss2}
\end{equation}
This is obtained by replacing $N^{-1/\nubar}$ with $L = \log_2 N$ 
in eq.~(\ref{eq:fss1}). 
Note that $L$ is locally approximated 
by a power function $N^{1/\nubar_\mrm{local}(L)}$ 
with $\nubar_\mrm{local}(L) = d \ln L / d\ln N = L \ln 2$, 
to reproduce eq.~(\ref{eq:fss1}) in a narrow range of $L$. 
The two scalings are compared in Fig.~\ref{fig:fss}. 
While the  scaling with $L$ shows good collapsing of data, 
the scaling with $N^{1/\nubar}$ breaks down for large $L$ 
(We use $1/\nubar=0.12$ in \cite{Baek09b} 
and treat larger generations by 7 than that in \cite{Baek09b}) 
and only works in the narrow size range, around $L=12$, 
as predicted from $1/\nubar_\mrm{local}(12) \approx 0.120$. 
Note that the scaling with $L$ works 
for any $p_b \in (p_{c1}, p_{c2})$ if $\phi$ equals $\psi(p_b)$ 
(numerically confirmed too, not shown here) 
and therefore it does not gives the threshold of the second transition. 
Presumably some irrelevant finite size effect 
or short range behavior of $C_0$ 
yields the best FSS fitting point $p_b$ 
which depends on the data range of $L$.

Another evidence for $p_b \approx 0.48$ shown in \cite{Baek09b} 
is the crossing of the ratio of the second largest cluster 
to the largest cluster, $\sratio$. 
Why crossing point gives critical point 
is based on the fact that 
the ratio $\sratio$ in the large size limit 
behaves as a step function of $p$ 
around the critical point 
and takes a special value in the middle of step 
on the critical point, 
which is clearly confirmed 
by the FSS in the square lattice in \cite{Baek09a}.
Again it is not clear 
whether this is also true for the transition of the EBT. 
If the critical {\it point} between the non-percolating and percolating phases 
is replace by the critical {\it phase}, 
characterized by fractional $\psi(p)$, 
it is naturally expected 
that a slope appears to fill the gap.
Such a slope is actually observed in the Cayley tree 
for $p_{c1} < p < p_{c2} = 1$ in \cite{Baek09a}. 
Indeed we observe a tendency in the large $L$ limit
that $\sratio$ converges to a value 
which continuously decreases for $p_{c1} < p < p_{c2}$ 
rather than forms a step at $p_b$ (not shown here). 
In addition, we confirmed that 
$\sratio$ is far from a universal function of $(p-p_b) N^{1/\nubar}$ 
(not shown here) unlike for the case of square lattice \cite{Baek09a}. 
The crossing of $\sratio$ 
is considered to be caused by 
the change of the tendency in irrelevant finite size effect.

In conclusion, 
we provided a simple scenario of 
the second percolation transition on the EBT 
and some numerical evidences which supports the scenario. 
We also showed that the FSS performed by Baek \etal 
does not holds for wide range of system sizes. 
Let us emphasize that the transitions of nonamenable graphs including the EBT 
is quite different from the usual second order transitions 
and standard analysis of second order transitions 
in amenable graphs cannot be applied directly to them. 
The value of $p_{c2}$ is, at least, larger than their estimation 
and the duality relation, $p_{c2} = 1-\pbar_{c2}$, 
seems valid for the percolation on the EBT. 
Baek \etal also claimed that 
the duality relation breaks down 
between the pair of \{3,7\} and \{7,3\} hyperbolic lattices 
based on the FSS analysis \cite{Baek09b}. 
We also consider they underestimate 
the second threshold probability in this model. 
The duality relation should be true in this model 
since both of the dual hyperbolic lattices 
are {\it transitive} in the large size limit \cite{Benjamini00}.


\section*{References}
\begin{thebibliography}{10}

\bibitem{Nogawa09} Nogawa T and Hasegawa T 
2009 J. Phys. A: Math. Theor. {\bf 42} 1450001

\bibitem{Benjamini00} Benjamini I and Schramm O 
2000 J. Am. Math. Soc. {\bf 14} 487

\bibitem{Baek09b} Baek S K, Minnhagen P and Kim B J 
2009 appeas in J. Phys. A: Math. Theor. ,
arXiv:0910.4340

\bibitem{Baek09a} Baek S K, Minnhagen P and Kim B J 
2009 Phys. Rev. E {\bf 79} 011124




\end{thebibliography}
\end{document}